\documentclass[prb,reprint,amsmath,amssymb,showpacs,superscriptaddress,floatfix]{revtex4-1}
\usepackage[breaklinks=true,colorlinks,citecolor=blue,linkcolor=blue,urlcolor=blue]{hyperref}
\usepackage{floatrow}

\usepackage{epsfig,mathrsfs,color,latexsym,subfigure,marginnote}
\renewcommand{\BibitemShut}[1]{}

\usepackage{csquotes}

\begin{document}
\title{First-principles cluster expansion study of functionalization of black phosphorene via fluorination and oxidation}

\author{Suhas Nahas}
\affiliation{Dept. of Material Science and Engineering, Indian Institute of Technology Kanpur, Kanpur 208016, India}
\author{Barun Ghosh}
\affiliation{Department of Physics, Indian Institute of Technology Kanpur, Kanpur 208016, India}
\author{Somnath Bhowmick}\email[]{bsomnath@iitk.ac.in}
\affiliation{Dept. of Material Science and Engineering, Indian Institute of Technology Kanpur, Kanpur 208016, India}
\author{Amit Agarwal}
\email{amitag@iitk.ac.in}
\affiliation{Department of Physics, Indian Institute of Technology Kanpur, Kanpur 208016, India}

\date{\today}
\begin{abstract}
Predicting the ground states for surface adsorption is a challenging problem because the number of degrees of freedom involved in the process is very high. Most of the studies deal with some specific arrangements of adsorbates on a given surface, but very few of them actually attempt to find the ground states for different adatom coverage. In this work, we show the effectiveness of cluster expansion method to predict the ``ground states'' resulting from chemisorption of oxygen and fluorine atom on the surface of monolayer black phosphorus or phosphorene. For device applications, we find that in addition to bandgap tuning, controlled chemisorption can change the unique anisotropic carrier effective mass for both the electrons and holes and even rotate them by 90$^\circ$, which can be useful for exploring unusual quantum Hall effect and novel electronic devices based on phosphorene. 
\end{abstract}
\maketitle
\section{Introduction}
Black phosphorus has emerged as a very promising 2D material, on account of its direct band gap, high carrier mobility, thickness dependent  and highly anisotropic electronic properties. \cite{PNAS_review,C4CS00257A, Li1, Liu14, Xia, Qiao1, Fei14, blackp,abbas2015,ge2015,jing2015,xu2015,biy2015} Moreover, few layers of black phosphorus has also been successfully exfoliated from its bulk counterpart and is available in free standing form.~\cite{PNAS_review,C4CS00257A, Li1, Liu14} As a shortcoming, due to the presence of a lone pair of electrons, phosphorus is highly reactive at the surface, and this can lead to the degradation of phosphorene (for example, via photo oxidation~\cite{utt2015}) under ambient conditions,\cite{Konig, degradation} hindering its potential applications. A possible solution to this problem may be chemical passivation of phosphorene with adsorption of monovalent and divalent atoms such as fluorine and oxygen. Apart from controlled passivation of phosphorene, chemisorption may also lead to new stable structures with interesting electronic properties as in the case of graphene, silicene and MoS$_2$.~\cite{PhysRevB.68.085410, Huang14, Huang13, rastogi2014} 

As observed by~\citeauthor{Odefects}~\cite{Odefects} based on first principles calculations, chemisorption of a \textit{single oxygen atom} at the phosphorene surface is an exothermic process. Interestingly, dangling or surface location (O bonded to only one P atom) is energetically more favorable than the bridge  or interstitial position (O bonded to two P atoms).\cite{Odefects} Such a defect, where O is dangling, causes minimal lattice distortion and it is also electrically neutral with no midgap states.\cite{Odefects,oxidation_Wang} Not only the bridge position is energetically less favorable because of large lattice distortion, but also it requires 0.7 eV of activation energy for one O atom to hop from a dangling to a bridge location.\cite{Odefects} In case of high O concentration (upto 250\%), various planar and tubular forms of oxides have been proposed with the former being energetically more favorable.\cite{PO} For the case of monovalent atom adsorption, first principles calculations predict better stability for fluorinated phosphorene as compared to hydrogenated phosphorene, which is found to be dynamically unstable.\cite{Boukhvalov}

One important but unexplored problem is to find the ground states among the many possible arrangements of adsorbate on phosphorene for a given adatom concentration. Since it involves sampling of a $2^N$ dimensional configuration space ($N$ is the number of lattice sites occupied by P atoms, each of which can either be coupled to an adatom or remain vacant), a ground state search by exhausting all the possible combinations by means for first principles calculations is not feasible. As a feasible alternative, we choose cluster expansion (CE) technique for searching the ground state at various different concentration in between pristine and fully oxidized/fluorinated phosphorene (PO/PF). Unlike mean field approximation, the CE technique is well known for giving a complete microscopic description of the system.\cite{Sanchez84,Ferreira89, Laks92, CET}

In this article, we predict several new ``ground states'' of fluorinated and oxidized phosphorene and also study the effect of adsorption on electronic properties with varying concentration of the adsorbate. Note that, we consider \textit{chemisorption only at the dangling position located at the surface} and these are possibly not the universal ground states, in particular at higher concentration of divalent oxygen, which can also be present at the interstitial bridge position. However, we argue later that such metastable states of oxidized phosphorene, predicted in this paper, can be stabilized kinetically and thus, possibly be synthesized under suitable experimental condition. Since fluorine is monovalent, it can only be adsorbed in the dangling position and we believe that the universal ground states are correctly predicted in this work. In terms of electronic properties, we find that other than bandgap modification, systematic functionalization of phosphorene can tune the anisotropy of {\it both} the electron and hole effective masses in a controlled way; unlike the case of external strain,  which can tune {\it only} the electron effective mass.~\cite{Fei14,baisheng2015} We also check the thermal stability of all the ``ground state'' structures by doing a molecular dynamics (MD) simulation. This article is organized as follows. In Sec.~II, we discuss the details of our computational approach. This is followed by a thorough discussion of fluorination of monolayer phosphorene in Sec.~III, and a discussion of oxidation of monolayer phosphorene in Sec.~IV. Thermal stability of discovered ``ground states'' are discussed in Sec.~V and  finally we summarize our results in Sec.~VI.

\section{Computational details}
\label{cd}
The ground state search at various different adatom concentrations is performed by using cluster expansion (CE) technique,~\cite{Sanchez84,Laks92,Ferreira89} as implemented in  alloy-theoretic automated toolkit (ATAT).\cite{van02,van02b,Connolly83} As the name suggests, the code can predict intermediate ground states of binary alloys, made by mixing different fractions of two pure components A and V. Surface adsorption can be treated using this method via a mapping  onto the pseudo-alloy problem by first identifying adatom locations (in the phosphorene unit cell), which can either be occupied by the adatom (component A) or remain vacant (component V). Thus, pristine and fully fluorinated/oxidized phosphorene corresponds to 100\% V -- 0\% A and vice versa, respectively. CE is a technique to parameterize the total energy for any given configuration of A and V for an arbitrary composition, expressed as A$_x$V$_{1-x}$ ($0<x<1$), thus avoiding the cumbersome task of computing total energy of 2$^N$ (N is of the order of lattice sites) structures from first principles calculations. As per the  CE formalism, any given configuration of A and V can be expressed by the vector $\boldsymbol{\sigma}=\{\sigma_1,\sigma_2,...,\sigma_n\}$, where $\sigma_i=+1$ or $-1$ depending on whether $i^{th}$ site is occupied by A (F/O) or V (vacant). Then, energy for a given configuration $\boldsymbol{\sigma}$ is expressed by an \enquote{Ising-like} Hamiltonian, 
\begin{equation} 
E(\boldsymbol{\sigma})=\sum\limits_\alpha m_\alpha J_{\alpha} \left\langle \prod_{i\in \beta} \sigma_i  \right \rangle
\end{equation}
where $\alpha$ represents a cluster, which is a collection of sites $i$, for example pairs, triplets, quadruplets etc. The multiplicities m$_\alpha$ is the number of clusters of type $\alpha$ and $J_\alpha$ is the corresponding effective cluster interaction (ECI). The sum is carried out over all clusters $\alpha$, which are not equivalent by space group symmetry and the ``spin'' product is averaged over all the clusters $\beta$, those are equivalent to $\alpha$-type clusters by symmetry. The ECIs ($J_\alpha$) are obtained by fitting the energies (of selected structures, generally 50-100 in number), obtained directly from {\it ab initio} calculations based on Density Functional Theory (DFT).\cite{Connolly83} The quality of the fit is determined by calculating the cross-validation score,\citep{van02}
\begin{equation}
CV^2=\frac{1}{n}\sum_{i=1}^n {(E_i^{DFT}-\hat{E}_i)}^2
\label{Eq_ham}
\end{equation}  
where $E_i^{DFT}$ is the DFT calculated energy of a particular structure $i$ and $\hat{E}_i$ is the predicted energy of the same structure, calculated from a least-square fit to the energies of the rest of the (n-1) structures (excluding $E_i^{DFT}$).

Once the energy is parameterized in terms of ECIs ($J_\alpha$), ground state for a particular adatom concentration $x$ is determined by comparing the formation energies, defined as
\begin{equation}
E_{\rm f} = E(A_xV_{1-x})-xE(A_1V_0)-(1-x)E(A_0V_1)
\label{Eq_for}
\end{equation} 
where $x$ denotes the adatom concentration and $E(A_1V_0)$ and $E(A_0V_1)$ are the energies of fully functionalized and pristine phosphorene, respectively. Relative stability of all the predicted ground states is compared in terms of binding energy, given by
\begin{equation}
E_b= \frac{1}{N_a}\left[E_{p+a}-(N_pE_p+N_{a}E_{a})\right]
\label{Eq_bin}
\end{equation}
where $N_a$ and $N_p$ denote the number of adatoms and phosphorus atoms in the unit cell, respectively. In Eq.~\eqref{Eq_bin}, $E_{p+a}$, $E_a$ and $E_p$  are the energies of adatom adsorbed phosphorene monolayer, a single phosphorus atom in pristine phosphorene monolayer and a free atom of fluorine or oxygen (depending on whether we consider fluorination or oxidation), respectively. 

The first principles (DFT) calculations are performed using a plane-wave basis set and ultrasoft pseudo-potential, as implemented in Quantum Espresso package.\cite{Giannozzi09} Electron exchange-correlation is treated with a generalized gradient approximation (GGA), using the formulation developed by Perdew-Burke-Ernzerhof (PBE).\cite{Perdew96} Kinetic energy cutoff for the plane-wave basis for wave functions is taken to be equal to 40 Ry. A well converged $k$-point mesh, depending on the size of each unit cell ($14\times 16\times 1$ for the smallest unit cell of pristine phosphorene), is chosen for the Brillouin-zone integrations. The structures are relaxed (both cell-vectors and ionic positions) until the the forces on each atom (total energy change due to ionic relaxation between two successive steps) is less than $10^{-5}$ Ry/au ($10^{-6}$ Ry). A supercell with a vacuum of $\sim$25 \AA~ perpendicular to the atomic plane is used to avoid the interaction with spurious replica images. 
Using these parameters, the in-plane and out-of-plane bond lengths of pristine monolayer phosphorene are found to be 2.23 and 2.26 \AA, respectively, which are in very good agreement with the values reported in the literature\cite{Qiao1}. 
Additionally we will see later that the binding energy of F/O atoms is very high, which clearly shows that these atoms are chemically bonded to the P atoms, as a consequence we do not need to  take into account van der Waals interactions for this problem. 
However, in case of multiple layers, it would be appropriate to include van der Waals interactions, which mainly affects the inter-layer spacing.

\begin{figure*}[ht]
\begin{center}
\includegraphics[width=.95 \linewidth]{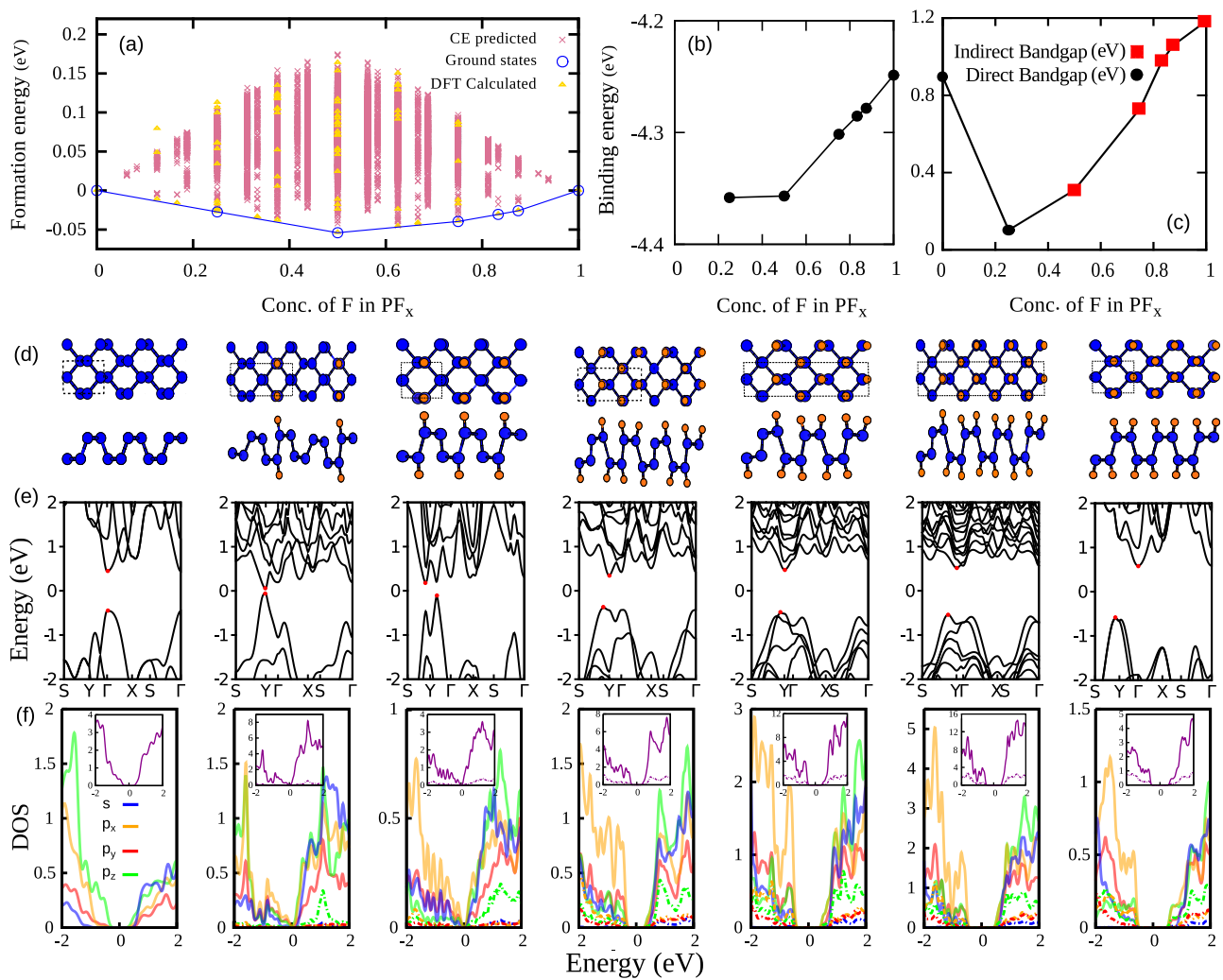}
\caption{(a) Formation energies $({\rm E}_{\rm f})$ at different fluorine concentration, calculated from DFT (triangles) and fitted values (cross) from Eq.~\ref{Eq_ham}. Blue circles indicate the ground states, obtained for $x=\{0,0.25,0.5,0.75,0.833,0.875,1.0\}$, whose binding  energy [calculated using Eq.~\eqref{Eq_bin}] and bandgap values are plotted in (b) and (c), respectively. (d) Unit cell (top and side view) of each of the ground states with increasing concentration of fluorine, and (e) the corresponding electronic band-structures. (f) The orbital resolved density of states (DOS) of the respective ground states, with the inset denoting the total DOS. Note that the solid and dashed lines correspond to phosphorus and fluorine, respectively.
\label{f1}}
\end{center}
\end{figure*}
 
\begin{figure}[t]
\begin{center}
\includegraphics[width= \linewidth]{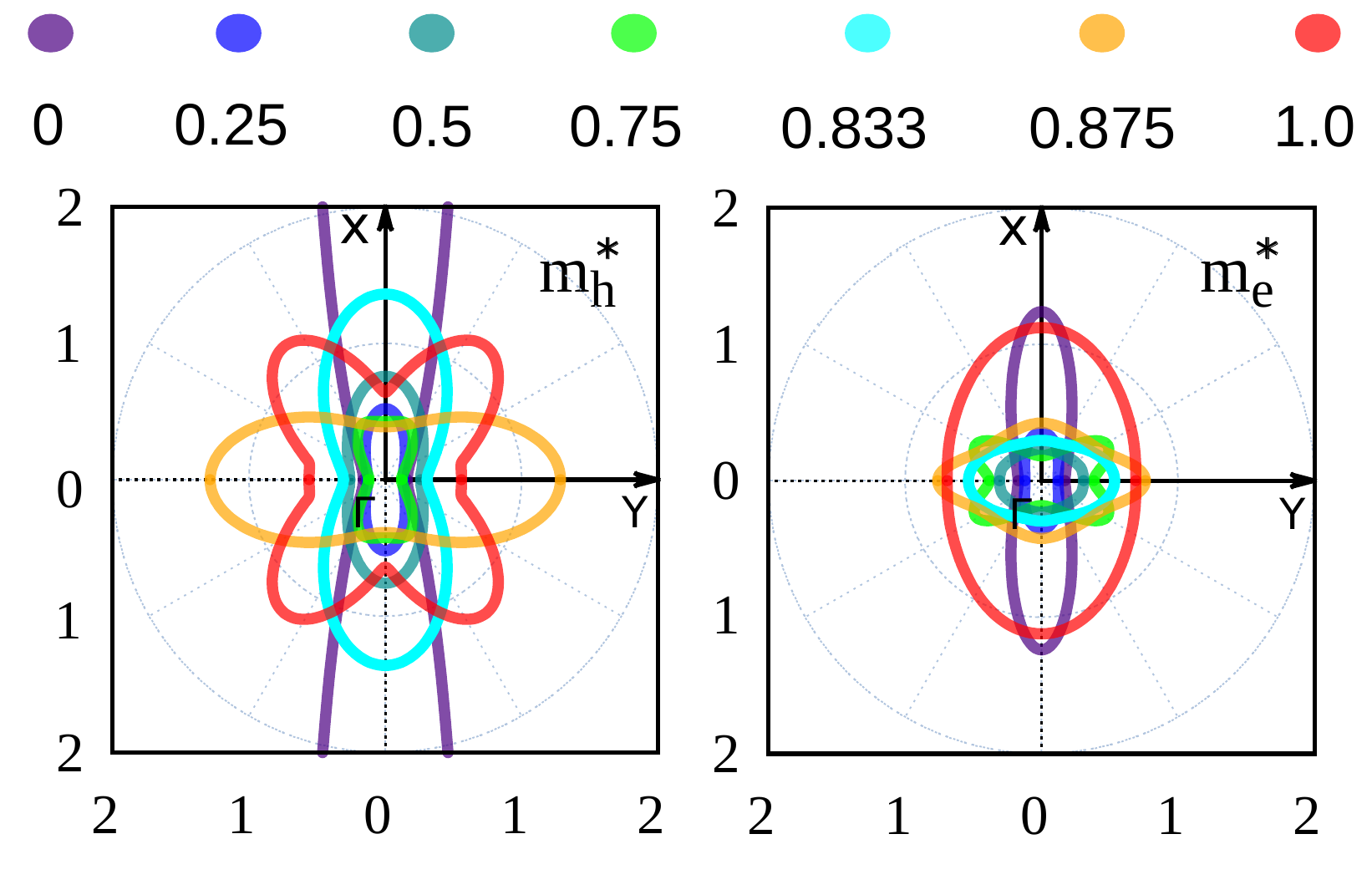}
\caption{ Polar plot of the hole and electron effective masses for various ground states of fluorinated phosphorene. Note that, m$_h^*$ has a two fold rotation symmetry for all the ground states, other than the fully fluorinated phosphorene, where the four fold rotation symmetry is evident. On the other hand, m$_e^*$ maintains it's two fold rotation symmetry, although becoming less anisotropic after complete fluorination.
\label{f5}}
\end{center}
\end{figure}

\section{Fluorination of Phosphorene}
\label{FP}
Unlike silicene, which is grown on a suitable substrate,~\cite{Lalmi10} monolayer phosphorene is a 2d-material that can be mechanically exfoliated~\cite{phosphorene1} and it is found to be stable in freestanding form, similar to graphene~\cite{graphene1} and MoS$_2$.~\cite{Radisavljevic11}  Thus, our ground state search includes both single sided as well as double sided functionalization. However, we find that the lowest energy structures are always those formed by double sided adsorption in case of fluorination, as well as oxidation. Fluorine is chosen as the monovalent adsorbate because hydrogenated structures of phosphorene are found to be unstable.~\cite{Boukhvalov} 

Applying cluster expansion technique, we first determine the ECI based on total energy calculation of 74 structures using DFT method. A cross validation score of 20 meV reflects that the difference between fitted and actual (from DFT calculations) energy of the fluorinated structures is negligible. Thus, based on the calculated ECI values, formation energies of 18652 fluorinated phosphorene structures (up to 32 atoms per unit-cell) are predicted and the true and predicted ground states agree very well [see Fig.~\ref{f1}(a)].
We find six ground states of fluorinated phosphorene: P$_{8}$F$_{2}$, P$_{4}$F$_{2}$,  P$_{8}$F$_{6}$, P$_{12}$F$_{10}$, P$_{16}$F$_{14}$ and P$_{4}$F$_{4}$  at specific fluorine concentrations of $x= 0.25, 0.5, 0.75, 0.833, 0.875$ and $1.0$, and the atomic arrangements, along with the unit cells are shown in Fig~\ref{f1}(d). The binding energy decreases with increasing fluorine concentration [see Fig~\ref{f1}(b)]. Note that, even at low F concentration the ground state configuration is constituted by two sided adsorption [see Fig~\ref{f1}(d)], which clearly indicates that one sided fluorination is energetically less favorable. A clear pattern is also visible, where the adatom located on a phosphorus atom in the upper plane of the puckered structure is accompanied by another adsorbed fluorine on the neighboring phosphorus atom in the lower plane. Fluorination leads to significant distortions in the structural parameters  of the underlying monolayer phosphorene, which in the pristine form has a bond-length of  2.2 {\AA} and bond-angles of  96$^{\circ}$  and 104$^{\circ}$. The structural parameter data for pristine phosphorene are in good agreement with the values reported in the literature.~\cite{wei2014} On fluorination, the bonds corresponding to the pair of phosphorus atoms to which the F atoms are adsorbed get elongated (bond-length ranging from  2.5 {\AA} to 3 {\AA}) while the bond-angles show small variations (up-to 5\%). The remaining P-P bonds, to which fluorine is not attached show negligible distortions.

\begin{figure*}[ht]
\begin{center}
\includegraphics[width=.95 \linewidth]{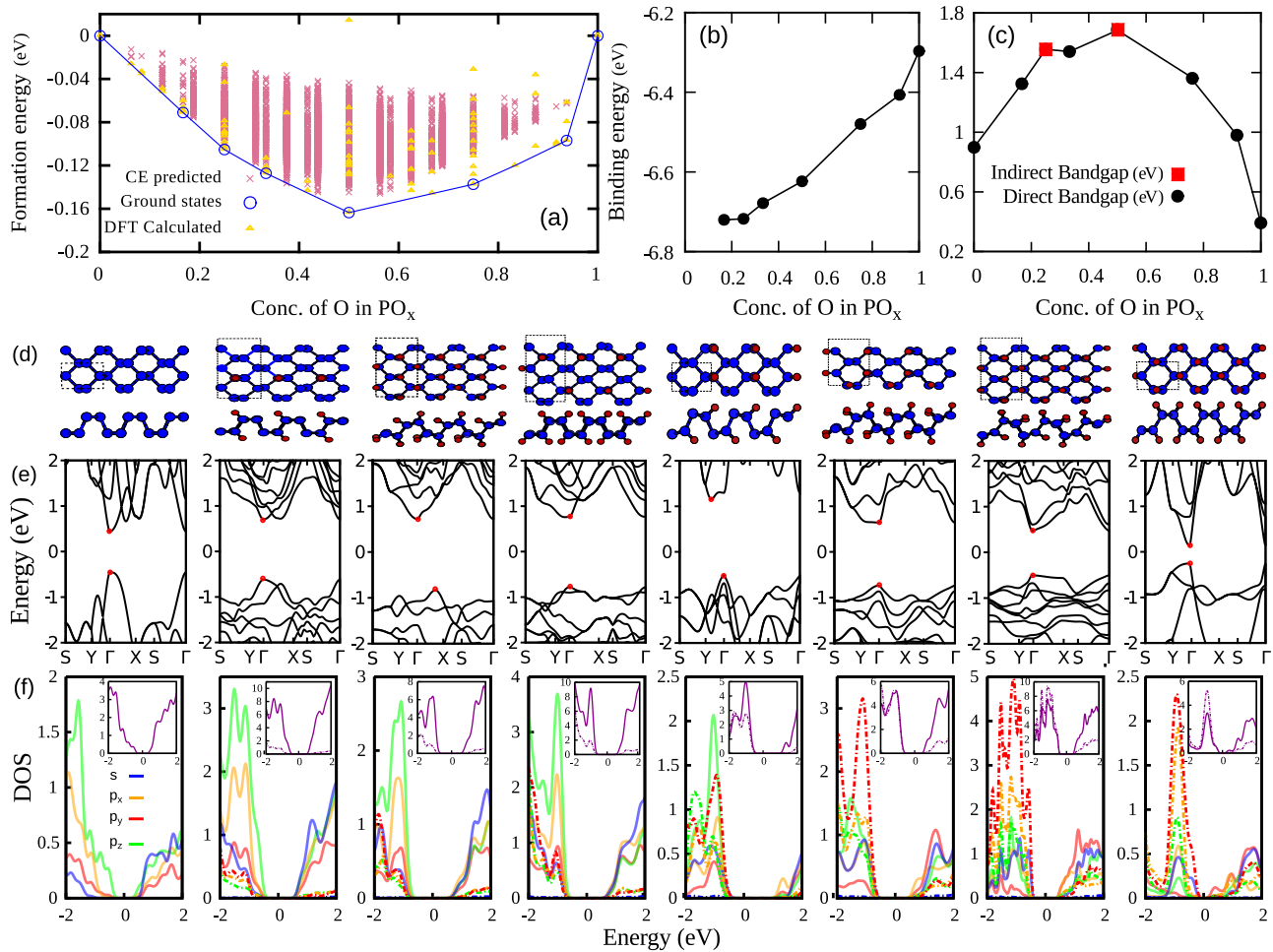}
\caption{(a) Formation energies $({\rm E}_{\rm f})$ at different oxygen concentration, calculated from DFT (triangles) and fitted values (cross) from Eq.~\ref{Eq_ham}. Blue circles indicate the ``ground states'', obtained for $x=\{0, 0.167, 0.25, 0.33, 0.5, 0.75, 0.917, 1.0\}$, whose binding  energy [calculated using Eq.~\eqref{Eq_bin}] and bandgap values are plotted in (b) and (c), respectively. (d) Unit cell (top and side view) of each of the ``ground states'' with increasing concentration of oxygen, and (e) the corresponding electronic band-structures. (f) The orbital resolved density of states (DOS) of the respective ``ground states'', with the inset denoting the total DOS. Note that the solid and dashed lines correspond to phosphorus and oxygen, respectively.
\label{f3}}
\end{center}
\end{figure*} 
   
Next we focus on electronic band structure, which changes significantly due to fluorine adsorption. Based on spin-resolved electronic structure calculations we find that fluorination does not induce any magnetization in phosphorene. Figure \ref{f1}(e) displays the electronic band structure of fluorinated phosphorene with increasing fluorine concentration. The energy dispersion relation is plotted along the high symmetry directions (S-Y, Y-$\Gamma$, $\Gamma$-X, X-S and S-$\Gamma$), where the high-symmetry points in the Brillouin zone are as follows: $\Gamma$(0,0,0), X(0.5,0,0), Y(0,0.5,0) and S (0.5,0.5,0). The direct band gap of pristine phosphorene (0.9 eV at $\Gamma$ point, within GGA) is initially found to decrease to 0.1 eV (Y point) at x=0.25. With further increase in fluorine concentration, the bandgap undergoes a transition from direct to indirect, accompanied by a monotonically increasing magnitude, reaching a value of 1.19 eV in fully fluorinated phosphorene (x=1). This variation of the bandgap as a function of fluorine concentration is shown in Fig.~\ref{f1}(c). Even though fluorination significantly alters the electronic structure of pristine phosphorene, both conduction band minima (CBM, corresponds to electron like charge carriers) and valence band maxima (VBM, corresponds to hole like charge carriers) states are still dominated by the orbitals belonging solely to the phosphorus atoms. Thus fluorine adatoms are not going to play any active role in carrier transport in black phosphorene. However, unlike pristine monolayer phosphorene where $p_z$ orbitals contributes most to the VBM states, it is the $p_x$ and $p_y$ orbitals which dominate in fluorinated phosphorene [see Fig.~\ref{f1}(f)]. In case of conduction band edge, $s$ as well as all the $p$ orbitals of phosphorus have significant weightage in pristine as well as fluorinated phosphorene.

One of the exciting properties of phosphorene is it's anisotropic electrical conductance and optical responses,~\cite{Liu14} which is attributed to the anisotropy of the effective mass of the charge carriers.~\cite{Fei14,baisheng2015} As shown in Fig.~\ref{f1}(e), compared to the $\Gamma$-Y direction, both the valence and conduction band in the $\Gamma$-X direction are relatively flat in pristine phosphorene near the $\Gamma$ point. Thus, both electron (m$_e^*$) and hole (m$_h^*$) effective mass is higher in the $\Gamma$-X direction (m$_e^*$=1.2 m$_e$ and m$_h^*$=8.64 m$_e$, where m$_e \approx 9.11\times 10^{-31}$ kg), than compared to their values in the $\Gamma$-Y direction (m$_e^*$=0.17 m$_e$ and m$_h^*$=0.16 m$_e$). The effective mass tensor is diagrammatically presented in Fig.~\ref{f5}, which approximately looks like the number 8 for pristine phosphorene. It has been shown that by applying strain, the electron effective mass tensor can be rotated by 90$^\circ$, taking the shape of \rotatebox{90}{8}.~\cite{Fei14,baisheng2015} In contrast, the hole effective mass is not so sensitive to the external strain; the nature of the anisotropy remains unchanged, with significantly higher effective mass in $\Gamma$-X direction than compared to $\Gamma$-Y direction.~\cite{Fei14,baisheng2015} From an application point of view,  this phenomenon can be used to switch the \textit{easy direction} of electronic and thermal transport in  phosphorene based devices. 

Since fluorination changes the location of VBM and CBM, accompanied by significant alteration of curvature of the respective bands [see Fig.~\ref{f1}(e)], we explore whether adatom adsorption can be an alternative way of tuning the anisotropic effective mass of phosphorene.  As shown in Fig.~\ref{f5}, m$_h^*$  in the $\Gamma$-X direction decreases significantly  from it's value of 8.64 m$_e$ in pristine phosphorene, lying in a range from 0.5 m$_e$ to 1.2 m$_e$ after fluorination. On the other hand, m$_h^*$ in $\Gamma$-Y direction does not change significantly after fluorination (ranging from 0.15 m$_e$ to 0.25 m$_e$, which is very close to it's value of 0.16 m$_e$, found in pristine phosphorene), except for high concentration of fluorine. As shown in Fig. \ref{f5}, value of m$_h^*$ in $\Gamma$-Y direction for x=0.875 and x=1.0 is 1.28 m$_e$ and 0.55 m$_e$, respectively. Interestingly, only for x=0.875, the anisotropy of m$_h^*$ is rotated by 90$^\circ$ and it's value in $\Gamma$-Y direction is nearly two times higher than that of $\Gamma$-X. In case of fully fluorinated phosphorene, the effective mass plot has a butterfly like shape. Broadly speaking, fluorination of phosphorene leads to significant decrease of m$_h^*$ in the $\Gamma$-X direction, while it is weakly affected in the $\Gamma$-Y direction, except at high adsorbate concentration. 
    
In case of electrons, effective mass in the $\Gamma$-X direction decreases due to fluorination, lying in a range from 0.1 m$_e$ to 0.3 m$_e$, other than the case of fully fluorinated phosphorene (1.1 m$_e$), where it is nearly equal to m$_e^*$ of pristine phosphorene (1.2 m$_e$). On the other hand, effective mass calculated along the $\Gamma$-Y direction increases moderately; after fluorination m$_e^*$ lies in the range from 0.25 m$_e$ to 0.75 m$_e$ except for the case of x=0.25, for which it is same as the value observed for pristine phosphorene (0.17 m$_e$). Note that, at higher fluorine concentration (x=0.833, 0.875), effective mass anisotropy is rotated by 90$^\circ$, such that m$_e^*$ in $\Gamma$-Y direction is almost two times larger than it's value along $\Gamma$-X. For x=1.0, effective mass plot takes the shape of an ellipse, with major axis (i.e., higher m$_e^*$) reverted to the $\Gamma$-X direction. In brief, due to fluorination m$_e^*$ calculated along the $\Gamma$-X direction decreases, while it increases in the $\Gamma$-Y direction. Rotation of the effective mass anisotropy by 90$^\circ$ is observed only for x=0.833 and 0.875.
 
\section{Oxidation of Phosphorene}
\label{OP}

Using cluster expansion method, at first we calculate the ECI values based on total energy calculation of 98 structures using DFT method. A cross validation score of 20 meV indicates that the difference between fitted and actual (from DFT calculations) energy is negligible. Next, using the calculated ECI values formation energies of 18650 phosphorene oxide structures (up to 32 atoms per unit-cell) are predicted [see Fig.~\ref{f3}(a)] and the true and predicted ``ground states'' match very well. In Fig.~\ref{f3}(d), we illustrate the atomic arrangement and unit cell of the seven predicted ``ground states'' of phosphorene oxide, namely P$_{12}$O$_{2}$, P$_{8}$O$_{2}$, P$_{12}$O$_{4}$, P$_{4}$O$_{2}$, P$_{8}$O$_{6}$, P$_{12}$O$_{11}$ and P$_{4}$O$_{4}$ for oxygen concentration of $x =$ {0.167, 0.25, 0.33, 0.5, 0.75, 0.917 and 1.0}, respectively. As shown in  Fig.~\ref{f3}(b), the binding energy decreases with increasing oxygen concentration. Note that, the oxygen adatoms are more strongly bound to monolayer phosphorene at any given concentration, compared to fluorine adatoms [see Fig.~\ref{f1}(b) and Fig.~\ref{f3}(b)]. Moreover, contrary to the case of fluorination, it is noticed that oxidation does not bring about significant changes in the P-P bond-lengths than compared to the pristine values. In fully oxygenated phosphorene, the longest P-P bond-length is found to be 2.38 {\AA}, which is 6.25 \% more than the pristine P-P bond-length. However, significant orientational distortions are observed with bond-angles ranging up to 125$^{\circ}$ for P-P-P triplets for x= 0.75 and 0.917.
 
A detailed study of electronic properties of the newly found ``ground states'' of phosphorene oxide reveals several interesting facts. Similar to the case of fluorination, first we rule out the possibility of any magnetic ground state of phosphorene oxide by doing spin-polarized electronic structure calculation. Next, we find that with increasing oxygen concentration, bandgap initially increases from 0.9 eV in pure pristine phosphorene to 1.68 eV at x=0.5, followed by a continuous decrease to 0.39 eV in fully oxidized phosphorene [see Fig.~\ref{f3} (c)]. The structure and the corresponding electronic band structure of all the predicted ``ground states'' of phosphorene oxide are shown in Figs.~\ref{f3} (d)-(e). As shown in the figure, most of the ``ground states'' of phosphorene oxide are direct bandgap semiconductors (other than x=0.33 and x=0.75), unlike indirect bandgap ground states of fluorinated phosphorene [see Fig.~\ref{f1} (c)]. It is clear from the PDOS plots [shown in Fig.~\ref{f3} (f)] that for $x<0.5$, both the conduction and valence band states are dominated by orbitals belonging to the phosphorus atoms. However, for $x>0.5$ the valence band states are prevailed by the orbitals belonging to the oxygen atoms, while the conduction band states are still dominated by the phosphorus orbitals. This is unlike fluorinated phosphorene, where both the valence and conduction band edges are dominated by the electronic orbitals of phosphorus at any given concentration [see Fig.~\ref{f1} (f)].

\begin{figure}[t]
\begin{center}
\includegraphics[width=\linewidth]{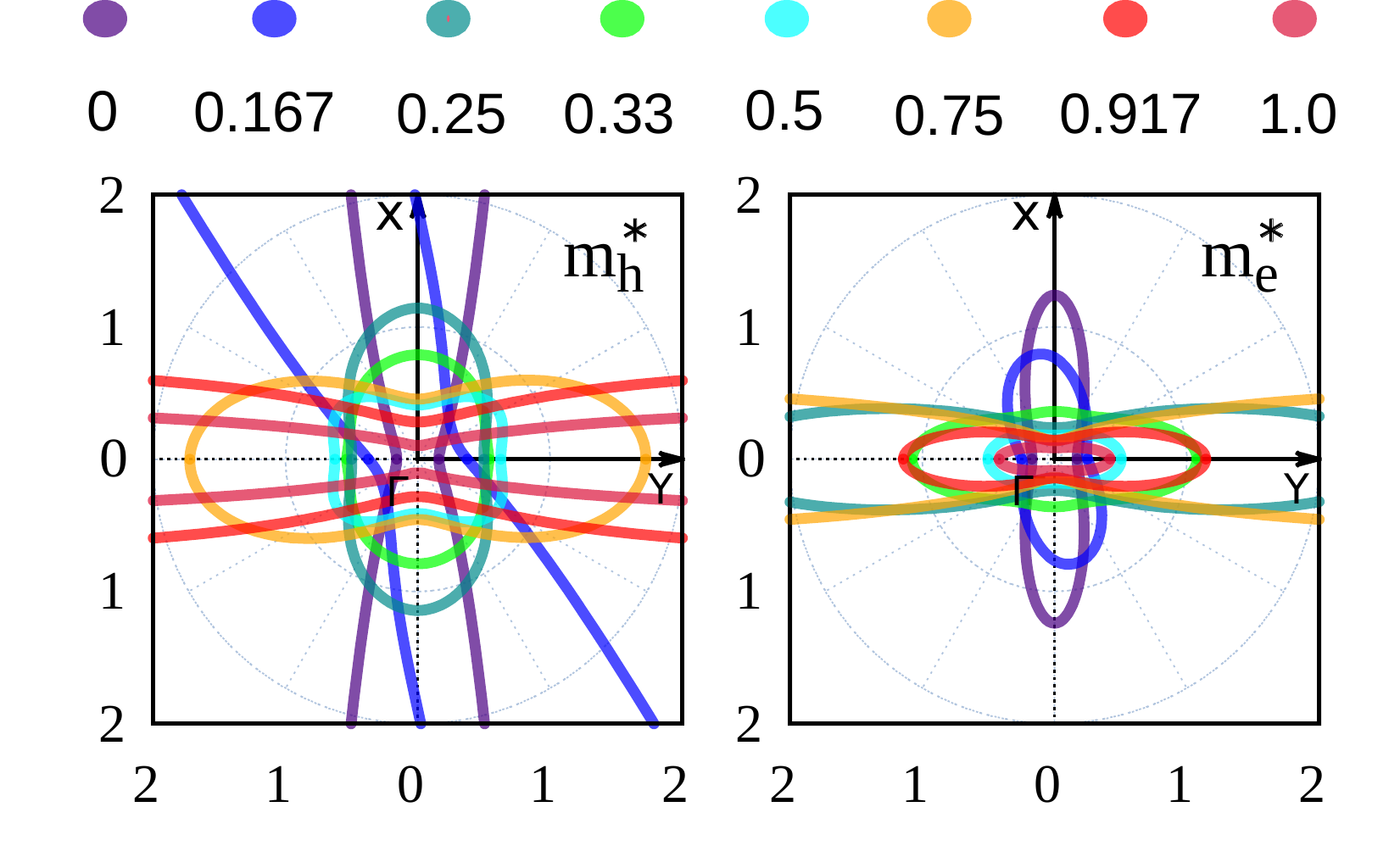}
\caption{Polar plot of the hole and electron effective mass for various ``ground states'' of oxidized phosphorene. Note that, while both m$_h^*$ and m$_e^*$ mostly maintains a two-fold rotation symmetry, the anisotropy of the effective mass undergoes a 90$^{\circ}$ rotation at higher oxygen concentration. 
\label{f6}}
\end{center}
\end{figure}

The hole effective mass shows a very interesting behavior after oxidation (see Fig.~\ref{f6}). In the $\Gamma$-X direction it decrease monotonically with increasing oxygen concentration; from 8.64 $m_e$ in pristine phosphorene to 0.10 $m_e$ in fully oxidized phosphorene. On the other hand, in the $\Gamma$-Y direction m$_h^*$ increases continuously from 0.16 m$_e$ in pristine phosphorene to 9.22 m$_e$ in fully oxygenated phosphorene. Similarly, the electron effective mass, calculated along the $\Gamma$-X direction, decreases monotonically with increasing oxygen concentration; from 1.28 $m_e$ in pristine phosphorene to 0.08 $m_e$ in fully oxidized phosphorene. On the contrary, in $\Gamma$-Y direction it increases from 0.17 $m_e$ in pristine phosphorene to 7.5 $m_e$ at $x=0.75$, beyond which it decreases continuously and reaches a value of 0.4 $m_e$ of fully oxidized phosphorene. Interestingly, both hole and electron effective mass contours undergo a 90$^\circ$ rotation after oxidation, which takes place for x$\ge$ 0.75 and x$\ge$ 0.25 for $m_h^*$ and $m_e^*$, respectively.

Thus, the unique anisotropic electron and hole effective mass of monolayer phosphorene can be modified in a controlled manner by adatom adsorption. This provides desirable tunability in terms of device applications, because spatial conductance profile of phosphorene can be engineered to undergo a  90$^{\circ}$ rotation of the preferred conduction direction of {\it both} electron and hole. Comparing Fig.~\ref{f5} and Fig.~\ref{f6}, we conclude that oxidation is more effective than fluorination for this purpose.  


Finally, let us discuss the choice of dangling vs. interstitial bridge position for the adsorption of oxygen atom. Since chemisorption in dangling location leads to minimal lattice distortion,\cite{Odefects} it is expected to be the favorable position for oxygen adsorption. Although this is found to be correct upto $\sim$25\% oxygen concentration, beyond which interstitial bridge position is predicted to be energetically more favorable.\cite{PO} Thus, some of ``ground states'' found for higher oxygen concentration in this paper are possibly metastable states. Interestingly, it is found that the activation barrier for the surface oxygen atom (at dangling position) to move to an interstitial site (at bridge position) is $\sim$0.7 eV.\cite{Odefects}  Because of this, we believe that the dangling configuration could be kinetically stabilized even when they are not preferred thermodynamically.

\section{Thermal Stability}
\label{therm} 
\begin{figure}[t]
\begin{center}
\includegraphics[width=0.8\linewidth]{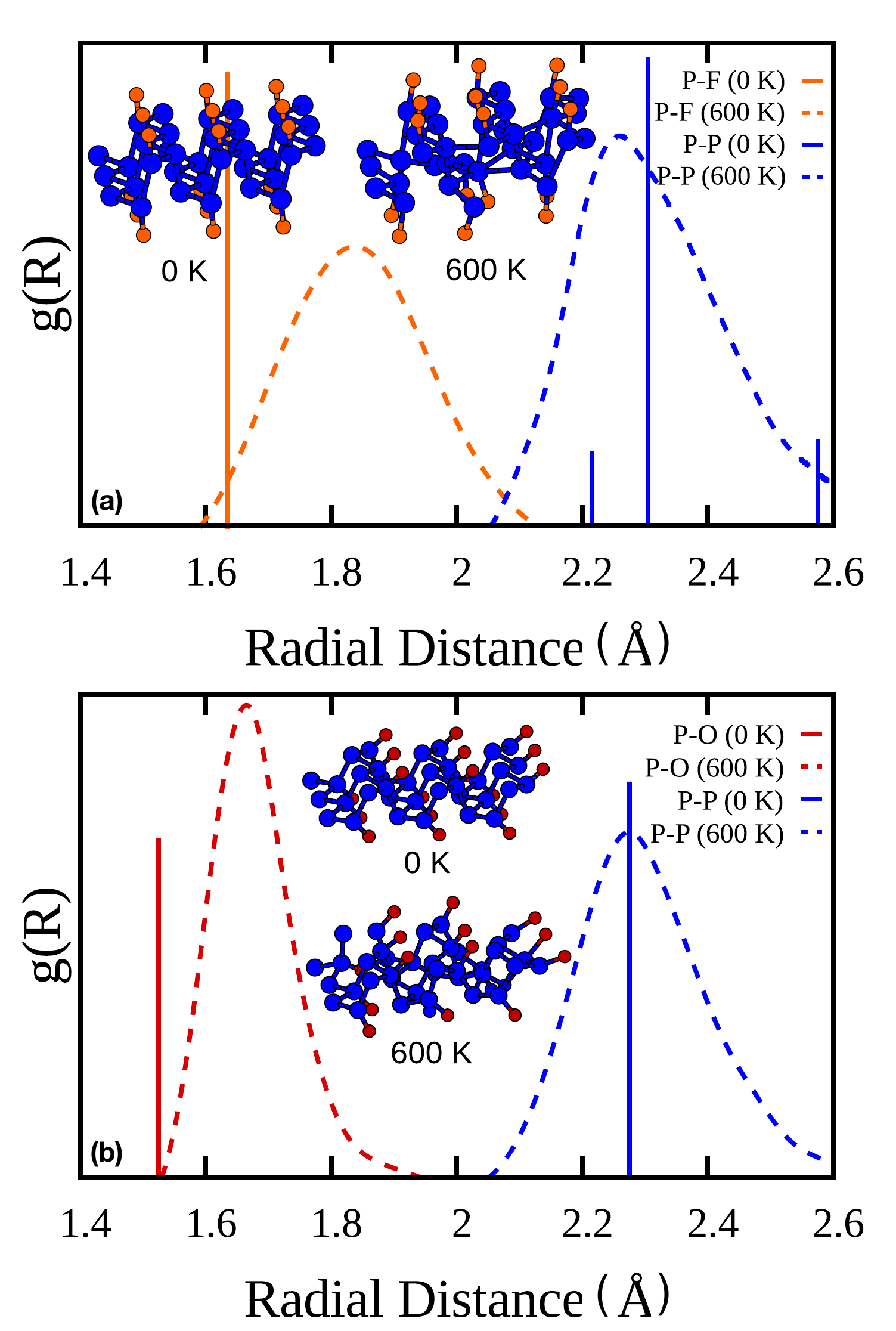}
\caption{The radial distribution function  $g(R)$ as a function of the radial distance $R$ for $T=0$ and $T=600$K for (a) PF$_{0.5}$ and (b) PO$_{0.5}$. Snapshots of the structures are also illustrated in the inset.  The broadening of the $T=0$ sharp lines at finite temperature ($600$ K), due to thermal vibrations of the molecules is evident.}
\label{f7}
\end{center}
\end{figure}
Before concluding, we confirm the stability of the predicted ``ground states'' at higher temperature by doing  Car-Parinello molecular dynamics~\cite{Car85} simulations, as implemented in the quantum espresso package.  Time step for MD simulation is taken to be 0.075 fs, with a total runtime of 1 ps and the fictitious electron mass ($\mu$)  is chosen to be 100 a.u to ensure that the electrons remain close to the ground state and  good control of the conserved quantities is maintained. All the reported ``ground states'' are found to be stable well beyond room temperature. We test the stability in the range 400--700 K by checking whether the structures disintegrate due to thermally induced vibrations. We find that most of the ``ground states'' are stable upto 700 K, other than PF$_x$ for $x>0.75$ coverage,  which are stable only upto 400--500 K temperature range.

Thermal stability for PF$_{0.5}$ and PO$_{0.5}$ is shown in terms of radial distribution function in Fig.~\ref{f7}(a) and (b), respectively. The sharp lines observed for 0 K are broadened at higher temperature due to thermally induced vibrations of the constituent atoms. We observe a peak broadening, measuring approximately $\pm 10\%$, which is very similar to the values reported for graphene.\cite{Fasolino09}

\section{Summary and Conclusion}
\label{Summary}
To summarize, we systematically study chemical functionalization of monolayer phosphorene by surface adsorption of fluorine and oxygen. Using CE method to analyze all possible structures (upto 32 atoms per unit cell), we have found the ground states at different concentrations of fluorine and oxygen, adsorbed on monolayer black phosphorene. It is observed that the bandgap \footnote{Magnitude of bandgap is known to be underestimated by 30-50\% due to GGA-PBE approximation. This can be corrected by hybrid functionals.\cite{hse06}}
as well as the effective mass of the charge carriers can be tuned by changing the concentration of adsorbates. In case of fluorination, most of the ground states are indirect bandgap semiconductors, while majority of the oxide ``ground states'' are direct bandgap semiconductors. Other than few selective adsorbate concentrations (x=0.25 and 0.5 for fluorination and x=1.0 for oxidation), the bandgap of the newly predicted ground states are either comparable or higher in magnitude than that of pristine phosphorene. Further we find that the anisotropy of both the electron and hole conductivity of phosphorene can also be tuned via adatom adsorption, which can lead to the rotation of the preferred conduction direction by 90$^{\circ}$. From an application point of view, this phenomenon can be used to control and switch the easy direction of electronic and thermal transport in phosphorene based devices.

\section{Acknowledgements}
A.A. acknowledges funding from the INSPIRE Faculty Award by DST (Govt. of India) and S.B. acknowledges funding from the SERB Fast-track Scheme for Young Scientist. Authors thank CC IITK for providing HPC facility. Crystal structures illustrated in this paper are drawn using XCrySDen software.\cite{xc1}  

\bibliography{ref}

\end{document}